# Pareto-Optimal Allocation of Transactive Energy at Market Equilibrium in Distribution Systems: A Constrained Vector Optimization Approach

Ahmad Khaled Zarabie, Sanjoy Das

**In a grid-constrained transactive distribution system market, distribution locational marginal pricing (DLMP) is influenced by the distance from the substation to an energy user, thereby causing households that are further away from the substation to be charged more. The Jain's index of fairness, which has been recently applied to alleviate this undesirable effect in efficient energy allocations, is used in this research to quantify fairness. It is shown that the Jain's index is strictly quasi-concave. A bilevel distributed mechanism is proposed, where at the lower level, auction mechanisms are invoked simultaneously at each aggregator to obtain energy costs under market equilibrium conditions. A constrained multi-gradient ascent algorithm, Augmented Lagrangian Multigradient Approach (ALMA), is proposed for implementation at the upper level to attain energy allocations that represent tradeoffs between efficiency and fairness. Theoretical issues pertaining to ALMA as a generic algorithm for constrained vector optimization are considered. It is shown that when the objectives are restricted to be strictly quasi-concave functions and if the feasible region is convex, ALMA converges towards global Pareto optimality. The overall effectiveness of the proposed approach is confirmed through a set of MATLAB simulations implemented on a modified IEEE 37-bus system platform.**

## I. NOMENCLATURE

| | |
|---|---|
| $\mathcal{N}$ | Set of nodes in grid, $|\mathcal{N}| = N$ |
| $\mathcal{A}$ | Subset of nodes with aggregators, $|\mathcal{A}| = A$ |
| $k$ | Node index, $k \in \mathcal{N}$ |
| $p_k$ | Power injection into node, $k \in \mathcal{A}$ |
| $c_k$ | Per unit cost of node $k \in \mathcal{A}$ |
| $\mathcal{G}_k$ | Set of agents at aggregator $k$, $k \in \mathcal{A}$, $G_k = |\mathcal{G}_k|$ |
| $\mathcal{G}_k^p, \mathcal{G}_k^c$ | Producer and consumer subsets of $\mathcal{G}_k$ |
| $i$ | Index of agent $i \in \mathcal{G}_k$ |
| $p_k^i$ | Energy demand of agent $i \in \mathcal{G}_k$ |
| $g_k^i, a_k^i, b_k^i$ | Generation and utility parameters of $i \in \mathcal{G}_k$ |
| $P_0$ | Total energy at DSO |
| $c_0$ | Unit cost at DSO |
| $\mathbf{p}_k$ | $G_k \times 1$ energy vector at aggregator $k \in \mathcal{A}$ |
| $\mathbf{c}$ | $A \times 1$ unit cost vector |
| $\mathbf{p}$ | $A \times 1$ energy allocation vector |
| $\mathbf{C}^V, \mathbf{c}_l^V, \mathbf{c}_u^V$ | Voltage constraint related constants |
| $\mathbf{C}^S, \mathbf{c}_0^S$ | Line capacity related constraint constants |
| $\mathbf{c}^{P_0}, c_0^{P_0}, \mathbf{c}_0^S$ | Energy balance condition related constants |
| $\Omega(\cdot)$ | Vector objective function |
| $\mathcal{W}(\cdot)$ | Global welfare function |
| $\mathcal{R}(\cdot)$ | Global fairness function |
| $\mathcal{J}(\cdot)$ | Jain's index of fairness |
| $\mathcal{U}_k(\cdot)$ | Aggregated utility of aggregator $k \in \mathcal{A}$ |
| $u_k^i(\cdot)$ | Utility of $i \in \mathcal{G}_k$ |
| $\mathcal{F}_\mathbf{x}$ | Feasible set of $\mathbf{x} \in \{\mathbf{p}, \boldsymbol{\omega}\}$ |
| $\underline{\boldsymbol{\alpha}}, \overline{\boldsymbol{\alpha}}, \boldsymbol{\beta}, \lambda, \gamma$ | KKT dual variables |
| $\boldsymbol{\xi}$ | Fritz-John dual variable |
| $\boldsymbol{\omega}$ | Feasible multi-gradient direction |
| $\nu$ | Scaling factor |
| $\eta_k^X$ | Step size with superscript $X$ denoting a variable |

In this list, only the main symbols used throughout the paper are shown. Other symbols are abstract combinations of these symbols or explained in the text. Bold lowercase and bold, italicized uppercase symbols represent vectors while bold uppercase symbols represent matrices. Scripted symbols are used to depict sets as well as global level functions.

## II. INTRODUCTION

DUE to recent advancements in technology as well as economic and environmental consideration, households may now be equipped with their own PV (photovoltaic) equipment, enabling them to participate in two-way energy exchange. Under these circumstances, uniform energy pricing policies are giving way to more elaborate differential pricing schemes. Distribution locational marginal pricing (DLMP) in transactive energy markets has received significant research attention [1],[2],[3],[4],[5],[6]. An unfortunate consequence of DLMP pricing is the inherent unfairness in charging users in the distribution grid, particularly those that are further away from the substation, with higher unit energy costs.

Energy allocation algorithms usually aim at maximizing the *welfare* i.e. total of all utilities of the prosumers (energy users, e.g. domestic households) in the grid [7] while ignoring the issue of fairness. Welfare maximizing algorithms are termed *efficient* algorithms in game theoretic parlance. Efficient approaches have routinely used in the energy market ([7], [8],[9],[10]). Published research on efficient energy allocation generally represent the transactive energy market with three categories of agents. The *distribution system operator* (DSO) is the uppermost agent. It acts as the interface between the market and external agencies such as the wholesale energy market. The DSO directly communicates with a set of *aggregator* agents, that are responsible for more real-time operation of the grid, as well as to decompose a DSO-level objective function into more computationally tractable components. Each aggregator communicates directly with its own set of prosumer agents, allowing them to participate in two-way energy trade.

In this research, a vector optimization algorithm is proposed to simultaneously maximize efficiency and fairness, the latter being quantified in terms of the Jain's index of fairness. The approach is an extension of the gradient ascent algorithm used in scalar optimization. The algorithm is capable of handling physical and other constraints imposed by the grid. Vector optimization is applied at the upper level of a bilevel framework and implemented by the DSO. The lower level incorporates distributed auction algorithms where energy users participate as bidding agents, is implemented by the aggregators.

*A. Related Work on Energy Markets*

Several DLMP-based pricing methods make use of DC optimal power flow (DCOPF) [1],[11],[12],[13] to establish physical constraints. Unfortunately DCOPF based approaches do not incorporate significant factors such as losses, voltage deviations, and reactive power flows essential in transactive distribution system markets. Furthermore, lower x/r ratios in the distribution system introduce significant errors in DCOPF-based DLMP [2]. Hence, a few recent approaches using some form of AC optimal power flow (ACOPF) have been proposed [2],[3],[4],[7]. Unfortunately, the issue of fairness has been largely overlooked.

Recently, a method in [14] applies the well-known Jain's index to alleviate the inherent spatial unfairness in DLMP approaches. This fairness measure is used in the form of a weighted regularization term that is added to a welfare-based objective function. The bilevel mechanism proposed therein obtains an otherwise efficient allocation of energy among the agents. Dual decomposition of the underlying constrained optimization problem, where the constraints are a result of grid's voltage and line flow limits and budget and energy balance conditions, provides the means for the DSO to directly compute DLMP costs. Since it is the DSO that determines (i.e. 'sets') the unit costs under DLMP, we call such a scheme a *cost-setting* mechanism.

A more game-theoretic scheme has been proposed recently where the DSO sets the energy allocated to the aggregators [7]. This *power-setting* mechanism allows unit costs to be established at the aggregator level using an auction process where the costs converge towards a fixed point. This scheme is illustrated in Fig. 1. where the DSO sends allocated power signals to the downstream aggregators, receiving unit costs from the latter. In cost-setting auctions, the signals would flow in the reverse direction; aggregators would receive costs from the DSO and send back their energy demands. Power-setting mechanisms offer a number of advantages over cost-setting ones, which are as follows.

(*i*) Power-setting mechanisms allow the prosumer agents to participate in auctions where they are allowed to buy or sell energy, in the same manner as other resources are traded in a classical marketplace.

(*ii*) Power-setting mechanisms establish Nash equilibrium among the set of participating agents in each aggregator. From a game-theoretic standpoint, this is highly significant as no prosumer agent stands to gain by consuming energy that deviates from its placed bid.

(*iii*) At the fixed points of the auctions, the unit costs are uniform over all agents and reflect market prices that automatically maximize the sum of all agent utilities within each aggregator – a phenomenon that is called "the invisible hand of the market" in Keynesian economics.

(*iv*) By shifting some of the computational burden to the aggregators, the DSO can use simpler optimization algorithms. For instance [7] uses proximal gradient ascent, which is not suitable for a cost-setting DSO.

(*v*) Similar market oriented approaches are gaining popularity in other engineering applications, such as spectrum allocation [15], cloud computing [16], and robot task allocation [17].

(*vi*) Power-setting auctions can easily operate during islanded grid conditions, where the distribution grid must function entirely in isolation.

The last observation warrants further elaboration. An islanded microgrid [18] is one that is isolated from upstream retailers. As the DSO does not receive any energy from external sources, transactions can only take place when PV-equipped households supply energy at higher costs to others. A cost-setting DSO that determines the monetary amount that downstream elements are charged in order to receive energy, is ill-equipped to operate in such situations. In sharp contrast, a power-setting DSO can easily establish energy trade using a market-oriented approach with the unit cost converging to the fixed point. In contrast to the cost-setting approach in [14], this research applies vector optimization under a power-setting format to exploit these advantages.

*B. Related Work on Vector Optimization*

Vector optimization are optimization approaches with vector objective functions. For decades, the simultaneous optimization of more than one objective function, has been implemented by means of multi-objective evolutionary algorithms such as genetic algorithms and particle swarm optimization (cf. [19],[20]). They have been applied to a plethora of problems in the energy market [21],[22],[23],[24],[25]. Multi-objective evolutionary algorithms are well equipped to handle constraints by either repairing infeasible solutions or simply rejecting them. Unfortunately as they incorporate stochastic operations, multi-objective evolutionary algorithms have to evaluate a large number of poor solutions (in terms of the objective functions) before converging to a Pareto-optimal set. As population-based approaches, evolutionary optimization cannot directly be applied to improve within a few quick steps any existing solution that is already close to Pareto optimality. Lastly, theoretical convergence guarantees of this class of algorithms are only of an indirect nature that approximate the algorithmic processes as discrete Markov chains. Lastly, evolutionary methods tend to be applied in situations without ascertaining the presence of a large number of local optima *a priori*, where simpler methods would have sufficed.

Deterministic vector optimization is a newer alternative to evolutionary algorithms. Normalized boundary intersection (NBI) is an indirect method of scalarization. It identifies the ideal solution in the objective function space from known theoretical bounds. Thereafter NBI directs the search towards the ideal point by means of conventional scalar optimization techniques. In [26], which applies NBI, the vector objective comprises of the scalar objectives of all aggregators. NBI is proposed as a method to minimize a vector of uncertainties in pricing in [27]. Unfortunately, NBI is prone to yielding solutions that are not in the Pareto front (i.e. the image manifold of the Pareto-optimal set). Conversely there are some regions in the Pareto front that are inaccessible to NBI [28].

Multi-gradient algorithms are a class of nascent algorithms that extend the steepest ascent method to vector objectives, some of them subsequently extended to quasi-Newton and second order methods. The early work in [29] defines a feasible ascent direction in terms of Hessian approximation. The *multi-gradient descent approach* (MGDA) in [30] uses a geometric definition of a feasible ascent direction. As MGDA is, to the best of the authors' knowledge, the only multi-gradient algorithm to have been adopted for any significant application domain [31],[32], it forms the basis of the approach proposed here. Unfortunately, MGDA is not equipped to handle constraints [33]. In [34], and more recently in [35], penalty function approaches are proposed to handle constraints. A very recent approach for constrained vector optimization (CVOP) has been proposed in [36] that extends Zoutendijk's method to handle active constraints.

It must be noted that there are numerous other approaches through which vector objectives can be handled, such as optimizing the weighted sums of objectives, lexicographic ordering, or elastic constraint methods [37]. The above discussion was confined only to the major classes of approaches that have found energy grid applications along with vector gradient ascent, which is relevant here.

*C. Technical Contributions*

This paper proposes a novel approach to obtain Pareto optimal energy allocations representative of the tradeoff between efficiency and fairness, where fairness is quantified in terms of the Jain's index. At first, necessary and sufficient conditions for Pareto optimality are formally established. The overall problem is formulated in terms of a CVOP, with the constraints comprising of the physical grid's voltage deviation and power flow limits, as well as cost and energy balance conditions. This vector objective framework allows a two-stage optimization algorithm to obtain tradeoff energy allocations that compromise some efficiency for more fair allocation.

The proposed algorithm, referred to hereafter as *augmented Lagrangian multi-gradient ascent* (ALMA), that is used to attain Pareto optimality, is implemented by the DSO. This bilevel approach applies primal decomposition to let the aggregators establish market equilibrium conditions, independently of one another and within their own subsets of energy users. Users in this research include seller agents in addition to a larger set of buyer agents. Due to DLMP pricing, it is imperative to keep track of dual variables in the iterative algorithm, which could not be implemented using the earlier approach in [7].

The major contributions of the proposed research are categorized below.

(*i*) The proposed ALMA relies on the recently proposed MGDA, which has so far only found limited applications. Furthermore, ALMA is a novel scheme that couples the well-known augmented Lagrangian method, which is intended for constrained scalar optimization, with an enhanced version of MGDA that can now handle constraints. To the best of the authors' knowledge, such an approach has not been used elsewhere in constrained vector optimization. Moreover, ALMA is a general-purpose approach for CVOP that can readily be adopted to similar engineering and other domains.

(*ii*) Although there is no dearth of literature on multi-objective optimization in energy systems, again to the best of the authors' knowledge, this research is the first to introduce to the energy systems community, an emergent class of vector optimization methods along with its accompanying mathematical underpinnings.

(*iii*) At the same time, the energy market offers the opportunity for ALMA's performance to be evaluated for large scale optimization, under the presence of a large number of constraints, that when put together outnumber the number of decision variables.

(*iv*) ALMA is implemented within a power-setting mechanism, thereby inheriting all the advantages proffered by the latter, as outlined earlier in Section I.A.

(*v*) The aggregator level auction used here is an improved version of that in [7],[38]. Agents no longer have to declare their intended roles as buyers or sellers beforehand; they can switch roles at any step based on changing energy costs during the auction.

(*vi*) As ALMA is built-in to this framework, off-the-shelf solvers are no longer required for optimization. This allows direct access to all quantities involved, including dual variables.

(*vii*) It is shown that Jain's fairness index is quasiconcave everywhere in the design space. Thereby, it extends previous results that showed its concavity only in the first orthant [6]. This is a significant result for any application that uses the index within an optimization procedure. Moreover, it is shown that the output of ALMA is Pareto-optimal as long as the objective functions are quasiconcave. Hence, ALMA is not restricted to purely concave utilities as in [4],[6],[7],[8],[13],[38].

(*viii*) To the extent of the authors' knowledge, this is the first attempt to seek Pareto-optimal efficiency-fairness tradeoffs with the Jain's index paired with the generic, widely accepted measure of welfare, instead of more synthetic measures of utilities that are specific to the Jain's index, i.e. $\alpha$-fair utility function [39],[40].

The rest of this paper is organized in the following manner. In Section III, the lower level auction algorithm is briefly introduced and the upper level CVOP, formulated. Section IV provides a theoretical analysis of ALMA. The simulation results are detailed in Section V. Finally this research concludes in Section VI where limitations of this approach are outlined, and future extensions suggested.

III. FRAMEWORK

Fig. 1 is a schematic of the bilevel framework. The upper level mechanism is implemented by the DSO, which possesses physical information pertaining to the distribution grid. It communicates power allocations $p_k$ from each aggregator, $k \in \mathcal{A}$, and receives equilibrium unit costs $c_k$ from them. Only a subset $\mathcal{A}$ of $\mathcal{N}$ are aggregators. Each aggregator $k$ contains a set $\mathcal{G}_k$ of prosumers within a physical neighborhood. The information flow between an aggregator and its agents $i \in \mathcal{G}_k$ are energy allocations, $p_k^i$ as well as unit costs, $c_k^i$.

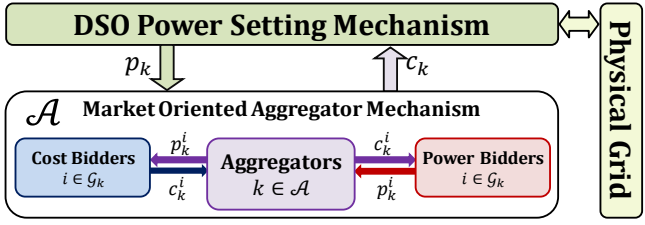

Fig. 1. Schematic of market-driven bilevel mechanism

*A. Aggregator Mechanism*

There are two sets of agents in each aggregator $k$ – the set $\mathcal{G}_k^P$ of power bidders that receive unit costs and return power bids, as well as the set of cost bidders $\mathcal{G}_k^C$ that receive power allocations and return cost bids. Both sets contain selfish agents that place bids to maximize their own payoffs, $u_k^i(p_k^i + g_k^i) - c_k^i p_k^i$. As shown in Fig. 2, the utility functions $u_k^i(\cdot)$ are assumed to be strictly quasiconcave, monotonically increasing, differentiable, and includes the origin (as in [41]). The quantity $g_k^i$ is the PV generation. A similar assumption of quasiconcave utilities has been adopted in [9] for energy trade between vehicles and the grid.

The auction algorithm is outlined below. The quantity $S_k$ is called the supply and is the sum of the power $p_k$ supplied by the DSO and those that the agents in $\mathcal{G}_k^P$ are willing to sell to the aggregator at unit cost $c_k$. Similarly, the quantity $R_k$ is the total monetary revenue that the aggregator will garner from the buying agents in $\mathcal{G}_k^C$. Although these can be initialized in various possible ways, e.g. randomly, the number of iterations can be reduced drastically if the converged values from a previous auction (of aggregator $k$) are used. The step where they as well as $p_k^i, c_k^i, A_k$ and $R_k$ are initialized is excluded from the outline of the auction.

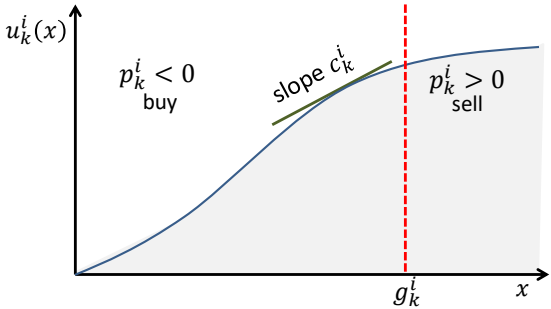

Fig. 2. Typical quasiconcave utility function of agent $i \in \mathcal{G}_k$

At the beginning of each iteration, the aggregator reassigns to $\mathcal{G}_k^C$ any agent $i$ that was previously in $\mathcal{G}_k^P$ but intends to buy power $p_k^i \geq 0$ (step 1). Likewise, it transfers any agent $i \in \mathcal{G}_k^P$ that has placed a unit cost bid $c_k^i < c_k$ to $\mathcal{G}_k^C$ (step 2). In step 3, the aggregator level unit cost $c_k$ is computed anew as the ratio of revenue $R_k$, to supply $S_k$. Next, the aggregator holds an auction within all sellers in $\mathcal{G}_k^P$ (step 4) and receives as bids, the amounts of power $p_k^i$ that they are willing to sell at the uniform rate, $c_k$. Following the sellers' auction, it updates the value of $A_k$. In step 6, which is referred to as *proportionally fair allocation* [7],[38], the aggregator divides the total power supply available $S_k$ among the agents in $\mathcal{G}_k^C$ in proportion to the total monetary amount that they are willing to pay. It holds an auction with the agents in $\mathcal{G}_k^C$ bidding new values of $c_k^i$ (step 7) In step 8, the aggregator updates $R_k$ using the received bids, by summing the products $c_k^i p_k^i$ over all agents in $\mathcal{G}_k^C$.

receive $p_k$ from DSO
until (market equilibrium) do
1. $\mathcal{G}_k^P \leftarrow \mathcal{G}_k^P \setminus \{i \in \mathcal{G}_k^P | p_k^i \geq 0\}$
   $\mathcal{G}_k^C \leftarrow \mathcal{G}_k^C \cup \{i \in \mathcal{G}_k^P | p_k^i \geq 0\}$
2. $\mathcal{G}_k^C \leftarrow \mathcal{G}_k^C \setminus \{i \in \mathcal{G}_k^P | c_k^i < c_k\}$
   $\mathcal{G}_k^P \leftarrow \mathcal{G}_k^P \cup \{i \in \mathcal{G}_k^P | c_k^i < c_k\}$
3. $c_k \leftarrow \frac{R_k}{A_k}$
4. $\forall i \in \mathcal{G}_k^P: \quad p_k^i \leftarrow \underset{x}{\mathrm{argmax}}(u_k^i(x + g_k^i) - c_k x)$
5. $S_k \leftarrow p_k - \mathbf{1}_{|\mathcal{G}_k^P|}^T [p_k^i]_{i \in \mathcal{G}_k^P}$
6. $\forall i \in \mathcal{G}_k^C: \quad p_k^i \leftarrow \frac{c_k^i p_k^i}{R_k} S_k$
7. $\forall i \in \mathcal{G}_k^C: \quad c_k^i \leftarrow \underset{x}{\mathrm{argmax}}\left(u_k^i\left(\frac{S_k p_k^i}{R_k} x + g_k^i\right) - x p_k^i\right)$
8. $R_k \leftarrow \mathbf{1}_{|\mathcal{G}_k^C|}^T [p_k^i]_{i \in \mathcal{G}_k^C} \circ [c_k^i]_{i \in \mathcal{G}_k^C}$

end do
send $c_k$ to DSO

As seen in steps 1 and 2 of each auction iteration, the agents are assigned as buyers or sellers using their bids from the preceding iteration. This is different from earlier versions of the algorithms in [7],[34]. Assuming that no agent is reassigned in steps 1 and 2, the aggregator mechanism's convergence towards a fixed point is shown in Fig. 3. It shows the supply $S_k$ of the sellers in response to a cost $c_k$ (red curve). The demand (blue curve) is the ratio $\frac{R_k}{S_k}$ after the buyers have placed their bids. Starting from an initial cost of $c_k = c_k^0$, the auction converges to $c_k = c_k^*$ in the counter-clockwise direction. It may be observed that a steeper supply curve would render the fixed point unstable. Divergence can be easily detected within two iterations of the auction, in which case the mechanism can be implemented in the clockwise direction for convergence. As this situation is unlikely to happen in any realistic setting (and in our simulations), it has not been elaborated further.

The aggregator's total utility is given by,

$$\mathcal{U}_k(p_k) = \mathbf{1}_{G_k}^T [u_k^i(p_k^i + g_k^i)]_{i \in \mathcal{G}_k} \quad (1)$$

Theorem-1 below shows that the aggregators reach an equilibrium unit cost $\mathbf{c}$ that is equal to $\nabla_\mathbf{p} \mathcal{W}(\mathbf{p})$, where $\mathcal{W}(\mathbf{p}) = \mathbf{1}_A^T [\mathcal{U}_k(p_k)]_{k \in \mathcal{A}}$ is the welfare. In [41] this gradient is referred to as the marginal benefit. It will be assumed that the agents do not bid strategically (see [25],[38]) so that the unit cost $c_k$ is independent of the placed bid $x = \{p_k^i, c_k^i\}$, i.e. $\frac{\partial c_k}{\partial x} = 0$. The proof of Theorem-1 is more straightforward than and distinct from the indirect one in [7], where the statement of Theorem-1 was shown to be a limiting case of virtual bidding.

**Theorem-1**. At the fixed point of the auction in aggregator $k$, the equilibrium cost is such that,

$$\nabla_\mathbf{p} \mathcal{W}(\mathbf{p}) = \mathbf{c}. \quad (2)$$

*Proof*: Consider a sellers' bidding strategy as shown in step 4. If the bid is $x$ its payoff is $u_k^i(x + g_k^i) - c_k x$. The payoff is maximum when its derivative with respect to $x$ is zero; so it places a power bid such that $\frac{\partial}{\partial x}\left[u_k^i(x+g_k^i)-c_k x\right]_{x=p_k^i}=0$, whence $u_k^{i'}(p_k^i+g_k^i)=c_k$.

Next, consider a buyer's bidding at any intermediate iteration. If the buyer responds to an allocation $p_k^i$ with a cost bid of $x$, from proportional allocation, its share of the total power will be $\frac{S_k p_k^i}{R_k}$ in the next iteration. Assuming a large $G_k$ such that $S_k$ and $R_k$ can be treated as constant with respect to the agent's bidding strategy, the bid is placed to maximize the overall payoff, $u_k^i\left(\frac{S_k p_k^i}{R_k}x+g_k^i\right)-x p_k^i$. This takes place in step 7. The derivative with respect to cost is $\frac{\partial}{\partial x}\left[u_k^i\left(\frac{S_k p_k^i}{R_k}x+g_k^i\right)-x p_k^i\right]_{x=c_k^i}$. Equated it to zero, we get, $\frac{S_k p_k^i}{R_k}u_k^{i'}\left(\frac{S_k p_k^i}{R_k}c_k^i+g_k^i\right)=p_k^i$. At the fixed point, in step 6 we must have, $p_k^i=\frac{c_k p_k^i}{R_k}S_k$, so that $c_k^i=\frac{R_k}{S_k}$. Under these circumstances the sellers bid is such that $u_k^{i'}(p_k^i+g_k^i)=c_k$.

Therefore it is seen that the equality $u_k^{i'}(p_k^i+g_k^i)=c_k$ applies to buyers and sellers. The statement of the theorem follows directly since $p_k=\mathbf{1}_{G_k}^T[p_k^i]_{i\in\mathcal{G}_k}$. ∎

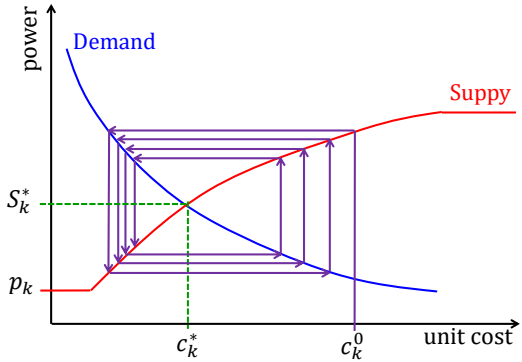

Fig. 3. Convergence towards fixed point of aggregator auction

### B. Constraints

Let $P_0$ be the total energy that the DSO receives from external sources at a unit cost $c_0$. The constraints imposed on ALMA are as follows. The voltages at all nodes in $\mathcal{N}$ must remain within their minimum and maximum limits (lower/upper voltage deviation constraints). The active and reactive power flows in the lines must not exceed their capacities (capacity limit constraint). Additionally, $P_0$ must equal the sum of the energy delivered to the aggregators and the losses occurring at the lines (energy balance condition). Lastly, the amount that the DSO must pay to external sellers must not exceed the total revenue obtained from the aggregator (budget balance condition). With appropriate values of all coefficients, these constraints can be expressed concisely as follows (detailed derivation can be found in [14]),

$$\begin{cases} -\mathbf{C}^V\mathbf{p}+\mathbf{c}_l^V\leq\mathbf{0}, & \text{(voltage deviation)} \\ \mathbf{C}^V\mathbf{p}+\mathbf{c}_u^V\leq\mathbf{0}, & \text{(voltage deviation)} \\ \mathbf{C}^S\mathbf{p}+\mathbf{c}_0^S\leq\mathbf{0}, & \text{(capacity limits)} \\ \mathbf{c}^{P_0 T}\mathbf{p}+c_0^{P_0}-P_0=0, & \text{(energy balance)} \\ -\mathbf{c}^T\mathbf{p}+c_0 P_0\leq 0, & \text{(budget balance)} \end{cases} \quad . \quad (3)$$

Thus, the feasible set of allocations $\mathbf{p}$ is given by,

$$\mathcal{F}_\mathbf{p}\triangleq\left\{\mathbf{p}\left|\begin{array}{c} -\mathbf{C}^V\mathbf{p}+\mathbf{c}_l^V\leq\mathbf{0},\mathbf{C}^V\mathbf{p}+\mathbf{c}_u^V\leq\mathbf{0}, \\ \mathbf{C}^S\mathbf{p}+\mathbf{c}_0^S\leq\mathbf{0}, \\ \mathbf{c}^{P_0 T}\mathbf{p}+c_0^{P_0}-P_0=0,-\mathbf{c}^T\mathbf{p}+c_0 P_0\leq 0 \end{array}\right.\right\}. \quad (4)$$

### C. Jain's Index of Fairness

The generic expression for Jain's index of fairness with argument $\mathbf{x}$ is as follows,

$$J(\mathbf{x})=\frac{1}{\|\mathbf{1}\|^2}\frac{(\mathbf{1}^T\mathbf{x})^2}{\mathbf{x}^T\mathbf{x}}. \quad (5)$$

The main motivation behind the choice of Jain's index as the measure of fairness is its *Schur concavity*, which is expressed as $\mathbf{x}\succcurlyeq\mathbf{y}\Rightarrow J(\mathbf{x})\geq J(\mathbf{y})$. In other words, if $\mathbf{x}$ *majorizes* $\mathbf{y}$ ($\mathbf{x}\succcurlyeq\mathbf{y}$), then $\mathbf{x}$ has a fairness index that is at least as high as that of $\mathbf{y}$. Majorization is explained as follows. Given the $n$ dimensional vector $\mathbf{x}$, let $\mathbf{x}_d^\dagger$ denote the $d<n$ dimensional vector formed by taking the numerically smallest $d$ elements of $\mathbf{x}$. For instance if $\mathbf{x}=[5\ 1\ 2\ 4\ 3]^T$ then $\mathbf{x}_3^\dagger=[1\ 2\ 3]^T$. The vector $\mathbf{y}_d^\dagger$ is obtained from $\mathbf{y}$ in an identical manner. We say that $\mathbf{x}$ majorizes $\mathbf{y}$ if and only if $\mathbf{1}_n^T\mathbf{x}=\mathbf{1}_n^T\mathbf{y}$ and $\mathbf{1}_d^T\mathbf{x}_d^\dagger\geq\mathbf{1}_d^T\mathbf{y}_d^\dagger,\ \forall d$. There is a more intuitive interpretation of this relationship, denoted as $\mathbf{x}\succcurlyeq\mathbf{y}$. Consider a pair of resource demand vectors, $\mathbf{x}$ and $\mathbf{y}$, with equal sums ($\mathbf{1}_n^T\mathbf{x}=\mathbf{1}_n^T\mathbf{y}$) allocated to $n$ aggregators. The quantities $\mathbf{1}_d^T\mathbf{x}_d^\dagger$ and $\mathbf{1}_d^T\mathbf{y}_d^\dagger$ are the sums of the resources received by the $d$ aggregators that have the least amount of resource allocated. Hence, $\mathbf{1}_d^T\mathbf{x}_d^\dagger\geq\mathbf{1}_d^T\mathbf{y}_d^\dagger$ implies that the resource-deprived aggregators collectively receive more resource through demand $\mathbf{x}$ than through demand $\mathbf{y}$. Hence, intuitively $\mathbf{x}\succcurlyeq\mathbf{y}$ means that $\mathbf{x}$ is fairer than $\mathbf{y}$.

In this research, the $G_k\times 1$ vector argument $\mathbf{x}$ of $J(\cdot)$ is determined as follows,

$$\mathbf{x}=\left[\frac{p_k}{c_k G_k}\right]_{k\in\mathcal{A}}. \quad (6)$$

Thus, each element consists of the energy $p_k$, normalized by the number of agents $G_k$ and the unit costs $c_k$. Dividing the power $p_k$ by $G_k$ ensures that each aggregator receives energy in proportion to the total number of household agents in it. The presence of $c_k$ in the denominator, is the DSO level version of proportional fairness [38], i.e. each aggregator allocated energy should be in proportion to the unit cost that the agents in it are willing to pay.

### D. Constrained Vector Optimization Problem Formulation

The welfare $\mathcal{W}(\mathbf{p})$ of the DSO is the sum of the utilities of all aggregators in $\mathcal{A}$. The fairness, expressed as a function of $\mathbf{p}=[p_k]_{k\in\mathcal{A}}$ is denoted as $\mathcal{R}(\mathbf{p})$, which is equal to $J(\mathbf{x})$ as shown in (5) where $\mathbf{x}$ is obtained from (6). The vector objective $\mathbf{\Omega}(\mathbf{p})$ that ALMA must simultaneously maximize with respect to $\mathbf{p}\in\mathcal{F}_\mathbf{p}$ where $\mathcal{F}_\mathbf{p}$ is as in (4), is given by,

$$\mathbf{\Omega}(\mathbf{p})\triangleq\begin{bmatrix}\mathcal{W}(\mathbf{p})\\\mathcal{R}(\mathbf{p})\end{bmatrix}. \quad (7)$$

Theoretical issues related to CVOPs as well as details of ALMA are discussed next.

## IV. PROPOSED APPROACH

### A. Theoretical Background

As mentioned earlier, ALMA although developed for this application, is a general-purpose algorithm for constrained vector optimization. Accordingly, as well as for conciseness, ALMA is discussed using more generic notation in Sections IV.A and IV.B. Without loss of generality, it is assumed that all objectives in the CVOP are to be maximized. Accordingly, let $\mathbf{f}: \mathbb{R}^n \to \mathbb{R}^m$ be the vector function to be maximized with respect to $\mathbf{x} \in \mathcal{F}_\mathbf{x}$, where $\mathcal{F}_\mathbf{x}$ is the feasible set,
$$\mathcal{F}_\mathbf{x} = \{\mathbf{x} | \mathbf{A}^\mathrm{T}\mathbf{x} + \mathbf{a} \leq \mathbf{0}, \mathbf{B}^\mathrm{T}\mathbf{x} + \mathbf{b} = \mathbf{0}\}. \quad (8)$$
Here, $\mathbf{A} \in \mathbb{R}^{n \times p}, \mathbf{a} \in \mathbb{R}^{p \times 1}, \mathbf{B} \in \mathbb{R}^{n \times q}, \mathbf{b} \in \mathbb{R}^{q \times 1}$. The feasible region $\mathcal{F}_\mathbf{p}$ introduced earlier in (4) clearly fits the generic form in (8). Scalar objectives in $\mathbf{f}(\cdot)$ are denoted as $f_j(\cdot), j \in \{1,2,\ldots,m\}$. It is assumed everywhere that each such function is Lipschitz continuous, and differentiable. Since in engineering optimization, the dimensionality of the design space is usually much higher than that of its image in the objective function space ($n \gg m$), it will be assumed hereafter that $n > m$.

Additionally, it will be assumed in this treatment, that the scalar objectives $f_j(\cdot)$ in $\mathbf{f}(\cdot)$ are *quasiconcave* functions. Quasiconcavity generalizes the notion of concavity. Any given function $f(\cdot)$ is (strictly) quasiconcave if and only if for every $\vartheta \in \mathbb{R}$, the upper contour set $\{\mathbf{x}|f(\mathbf{x}) \geq \vartheta\}$ is (strictly) convex. Two sufficient conditions for quasiconcavity are stated in the following axiom.

*Axiom*-1. With $\mathbf{x}, \mathbf{y} \in \mathcal{F}_\mathbf{x} \subseteq \mathbb{R}^n$ being any pair of vectors in the convex domain $\mathcal{F}_\mathbf{x}$, either of following inequality is a sufficient condition for the quasiconcavity of the function $f(\cdot)$,
$$\begin{cases} \theta \in [0,1] \Rightarrow f(\theta\mathbf{x} + (1-\theta)\mathbf{y}) \geq \min(f(\mathbf{x}), f(\mathbf{y})) = 0, \\ f(\mathbf{x}) \geq f(\mathbf{x} + \mathbf{y}) \Rightarrow \mathbf{y}^\mathrm{T}\nabla_\mathbf{x} f(\mathbf{x}) \geq 0. \end{cases} \quad (9)$$

The function's quasiconcavity becomes strict if all inequalities are strict in the first condition in (9). In addition to being sufficient, this condition is also a necessary one for quasiconcavity. It may be noted that the second condition in (9) defines pseudoconcave functions that is outside the scope of this research, and is therefore a stricter requirement than the first. Formal proofs can be found in [42],[43].

The relationship $\mathbf{f}(\mathbf{x}) > \mathbf{f}(\mathbf{y})$ is used to denote that $f_j(\mathbf{x}) > f_j(\mathbf{y}), \forall j$. Analogous elementwise interpretations apply to the remaining inequality relationships $\geq, <,$ and $\leq$. Given two vectors $\mathbf{x}, \mathbf{y} \in \mathcal{F}_\mathbf{x}$, $\mathbf{x}$ *weakly dominates* $\mathbf{y}$ when $\mathbf{f}(\mathbf{x}) \geq \mathbf{f}(\mathbf{y})$. When there is at least some $f_j(\cdot)$ such than $f_j(\mathbf{x}) > f_j(\mathbf{y})$, then $\mathbf{x}$ *dominates* $\mathbf{y}$. This (weak) dominance relationship is denoted as $\mathbf{x} \succ \mathbf{y}$ ($\mathbf{x} \succcurlyeq \mathbf{y}$). Any point[1] $\mathbf{x} \in \mathcal{F}_\mathbf{x}$ is *locally (weakly) Pareto-optimal* if and only if there exists a quantity $\sigma > 0$ satisfying the condition,
$$\mathbf{y} \in \mathcal{F}_\mathbf{x} \cap \mathcal{B}(\mathbf{x}, \sigma) \Rightarrow \mathbf{x} \succ \mathbf{y} \ (\mathbf{x} \succcurlyeq \mathbf{y}). \quad (10)$$
In the above expression, $\mathcal{B}(\mathbf{x}, \sigma) \subset \mathbb{R}^n$ is a ball centered around $\mathbf{x}$ with radius $\sigma$, $\mathcal{B}(\mathbf{x}, \sigma) = \{\mathbf{y} \in \mathbb{R}^n | \|\mathbf{y} - \mathbf{x}\| > \sigma\}$. If the condition (10) holds in the limiting case $\sigma \to \infty$, then $\mathbf{x}$ is said to be *(weakly) Pareto-optimal*. The image of the set of all Pareto-optimal points is the CVOP's Pareto front.

From here onwards, the $n \times m$ Jacobian matrix will be denoted as $\nabla_\mathbf{x}\mathbf{f}(\mathbf{x}) \triangleq [\nabla_\mathbf{x} f_1(\mathbf{x}) \cdots \nabla_\mathbf{x} f_M(\mathbf{x})]$. For simplicity it is assumed to be of full column rank unless noted otherwise. There is a useful relationship between the gradient vectors $\nabla_\mathbf{x} f_j(\mathbf{x})$ of the $j \in \{1,2,\ldots,m\}$ objectives of a locally Pareto optimal point $\mathbf{x}$. Consider another point, $\mathbf{y} = \mathbf{x} + \delta\mathbf{x} \in \mathcal{F}_\mathbf{x}$, where $\delta\mathbf{x}$ is an infinitesimal perturbation of $\mathbf{x}$ so that higher order terms in the Taylor's series expansion can be ignored. Hence, in the limiting case of $\delta\mathbf{x} \to \mathbf{0}$, $\mathbf{f}(\mathbf{y}) = \mathbf{f}(\mathbf{x}) + \nabla_\mathbf{x}^\mathrm{T}\mathbf{f}(\mathbf{x})\delta\mathbf{x}$. From (10), $\mathbf{f}(\mathbf{x}) > \mathbf{f}(\mathbf{y})$, so that $\nabla_\mathbf{x}^\mathrm{T}\mathbf{f}(\mathbf{x})\delta\mathbf{x} < \mathbf{0}$. Suppose $\delta\mathbf{x}$ is chosen such that all of its components are positive ($\delta\mathbf{x} > \mathbf{0}$). In other words, for every $f_j(\cdot)$ there must be at least one function $f_i(\cdot)$ such that $\nabla_\mathbf{x}^\mathrm{T} f_i(\mathbf{x})\delta\mathbf{x}$ and $\nabla_\mathbf{x}^\mathrm{T} f_j(\mathbf{x})\delta\mathbf{x}$ have opposite signs. This observation is significant. If such a $\delta\mathbf{x}$ does not exist, then from the convexity of $\mathcal{F}_\mathbf{x}$, another vector $\delta\mathbf{x} < \mathbf{0}$ exists, leading to the same observation. Another way of interpreting this observation is that from a Pareto optimal point any improvement (i.e. increase) in one objective can only be accomplished at the expense of another.

In CVOPs, there exist necessary and sufficient conditions that are analogous to the KKT conditions in scalar constrained optimization. These are the *Fritz-John* (FJ) *conditions* [42],[44],[45] for local Pareto optimality. We state these conditions in the following axiom.

*Axiom*-2. The vector $\mathbf{x} \in \mathcal{F}_\mathbf{x}$ is locally Pareto optimal if there exist vectors $\boldsymbol{\xi} \in \mathbb{R}^m$, $\boldsymbol{\lambda} \in \mathbb{R}^p$, and $\boldsymbol{\mu} \in \mathbb{R}^q$ satisfying the following conditions,
$$\begin{cases} \boldsymbol{\xi} \geq \mathbf{0}_m; \ \boldsymbol{\lambda} \geq \mathbf{0}_p \\ \boldsymbol{\lambda}^\mathrm{T}(\mathbf{A}^\mathrm{T}\mathbf{x} + \mathbf{a}) = 0; \\ \nabla_\mathbf{x}\mathbf{f}(\mathbf{x})\boldsymbol{\xi} - \mathbf{A}\boldsymbol{\lambda} - \mathbf{B}\boldsymbol{\mu} = \mathbf{0}_p. \end{cases} \quad (11)$$

The Fritz-John conditions reduce to the well-known first order KKT optimality conditions with $\boldsymbol{\xi} \neq \mathbf{0}_m$, the weighted sum of the objectives, $\boldsymbol{\xi}^\mathrm{T}\mathbf{f}(\mathbf{x})$ acting as the equivalent scalar objective and treating the function $\mathfrak{L}(\mathbf{x}, \boldsymbol{\xi}, \boldsymbol{\lambda}, \boldsymbol{\mu}) = \boldsymbol{\xi}^\mathrm{T}\mathbf{f}(\mathbf{x}) - \boldsymbol{\lambda}^\mathrm{T}(\mathbf{A}^\mathrm{T}\mathbf{x} + \mathbf{a}) - \boldsymbol{\mu}^\mathrm{T}(\mathbf{B}^\mathrm{T}\mathbf{x} + \mathbf{b})$ as the equivalent Lagrangian function, as seen in [32].

In multi-gradient ascent algorithms, the *common (feasible) ascent direction* is a vector $\boldsymbol{\omega}$ such that for some $\delta > 0$, $\mathbf{x} + \delta\boldsymbol{\omega} \succ \mathbf{x}$. Multi-gradient ascent involves iterative increments of $\mathbf{x}$ along common ascent directions. Using Taylor's series expansion it can readily be shown that any common ascent direction $\boldsymbol{\omega}$ must be expressed as a convex combination of the gradients,
$$\boldsymbol{\omega} = \nabla_\mathbf{x}\mathbf{f}(\mathbf{x})\boldsymbol{\xi}, \quad (12)$$
where $\boldsymbol{\xi} > \mathbf{0}$. In MGDA, the weights in $\boldsymbol{\xi}$ of the gradients $\nabla_\mathbf{x} f_j(\mathbf{x})$ are constrained so that $\mathbf{1}_m^\mathrm{T}\boldsymbol{\xi} = 1$. The direction $\boldsymbol{\omega}$ is chosen to be the minimum norm element in the convex hull of the gradients.

Axiom-3 below, stems from the observation made earlier that at a Pareto-optimal point, any gain with respect to an objective will always be at the expense of another. Formal proofs can be found in [30],[31],[46].

*Axiom*-3. At any point $\mathbf{x} \in \mathcal{F}_\mathbf{x}$, if no common feasible ascent direction satisfying (12) exists, then $\mathbf{x}$ is locally Pareto optimal.

*B. Augmented Lagrangian Multi-Gradient Ascent*

Since $\boldsymbol{\omega}$ in (12) will be used to increment $\mathbf{x} \in \mathcal{F}_\mathbf{x}$, we must have $\mathbf{x} + \boldsymbol{\omega} \in \mathcal{F}_\mathbf{x}$, so that, $\mathbf{A}^\mathrm{T}(\mathbf{x} + \boldsymbol{\omega}) + \mathbf{a} \leq \mathbf{0}$, and, $\mathbf{B}^\mathrm{T}(\mathbf{x} + \boldsymbol{\omega}) + \mathbf{b} = \mathbf{0}$. A *sufficient* condition on $\boldsymbol{\omega}$ to satisfy the above constraints would be that $\mathbf{A}^\mathrm{T}\boldsymbol{\omega} \leq \mathbf{0}, \mathbf{B}^\mathrm{T}\boldsymbol{\omega} = \mathbf{0}$. ALMA does not aim to bring the point $\mathbf{x}$ to the feasible region. This goal can be achieved separately through any other constrained optimization algorithm. The goal is merely that when ALMA increment **x** by **ω** should not violate the constraint any further. Inequality and equality constraints (Fig. 4) are considered separately below.

---
[1] The terms 'point' and 'vector' are used interchangeably.

(*i*) Suppose the inequality constraint is inactive so that $\mathbf{A}^T\mathbf{x} + \mathbf{a} = \boldsymbol{\delta}^a < \mathbf{0}$. In Fig. 4 (top), this corresponds to the small, green circle representing **x**. After replacing **x** with $\mathbf{x} + \boldsymbol{\omega}$, the constraint must not be violated. The analogous condition on $\mathbf{x} + \boldsymbol{\omega}$ is, $\mathbf{A}^T(\mathbf{x} + \boldsymbol{\omega}) + \mathbf{a} \leq \mathbf{0}$, which upon simplification yields, $\mathbf{A}^T\boldsymbol{\omega} \leq -\boldsymbol{\delta}^a$. Next, suppose the inequality constraint is violated so that $\mathbf{A}^T\mathbf{x} + \mathbf{a} = \boldsymbol{\delta}^a > \mathbf{0}$, with such an **x** shown as a red circle in Fig. 4 (top). For ALMA not to move $\mathbf{x} + \boldsymbol{\omega}$ further away from the feasible region than **x**, we must have $\mathbf{A}^T(\mathbf{x} + \boldsymbol{\omega}) + \mathbf{a} \leq \boldsymbol{\delta}^a$. In other words, $\mathbf{A}^T\boldsymbol{\omega} \leq \mathbf{0}$. Combining both the cases, the constraint upon the direction, **ω** must be $\mathbf{A}^T\boldsymbol{\omega} \leq -[\boldsymbol{\delta}^a]_-$. Here, $[\boldsymbol{\delta}]_- = \min(\boldsymbol{\delta}, \mathbf{0})$, with the minimization being carried out in a component-wise manner. Similarly, $[\boldsymbol{\delta}]_+ = \max(\boldsymbol{\delta}, \mathbf{0})$.

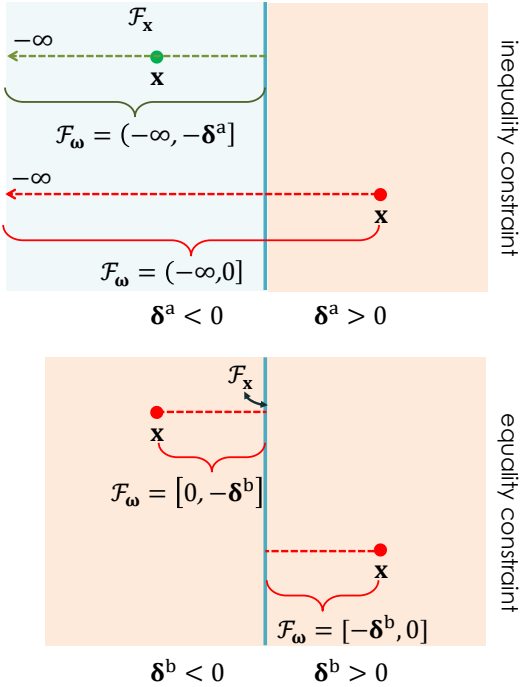

Fig. 4. Feasible regions of **ω**.

(*ii*) Suppose the equality constraint is violated in the positive direction so that $\mathbf{B}^T\mathbf{x} + \mathbf{b} = \boldsymbol{\delta}^b > \mathbf{0}$. Fig. 4 (bottom) depicts such a situation where the red circle is to the right of $\mathcal{F}_\mathbf{x}$ (vertical line). As we chose to ensure that the updated variable $\mathbf{x} + \boldsymbol{\omega}$ not move any further away from the feasible region, we must have, $\mathbf{0} \leq \mathbf{B}^T(\mathbf{x} + \boldsymbol{\omega}) + \mathbf{b} \leq \boldsymbol{\delta}^b$. This leads to the bounds, $-\boldsymbol{\delta}^b \leq \mathbf{B}^T\boldsymbol{\omega} \leq \mathbf{0}$. Next, suppose the equality constraint is violated in the negative direction. In this case, $\mathbf{B}^T\mathbf{x} + \mathbf{b} = \boldsymbol{\delta}^b < \mathbf{0}$ yielding the bounds, $\mathbf{0} \leq \mathbf{B}^T\boldsymbol{\omega} \leq -\boldsymbol{\delta}^b$. Combining both cases together, the equivalent condition that $\mathbf{x} + \boldsymbol{\omega}$ is no further away from $\mathcal{F}_\mathbf{x}$ than **x**, is given by, $-[\boldsymbol{\delta}^b]_+ \leq \mathbf{B}^T\boldsymbol{\omega} \leq -[\boldsymbol{\delta}^b]_-$.

Combining the observations for both kinds of constraints allows us to define feasible region $\mathcal{F}_\boldsymbol{\omega}$ for the direction **ω** in the following manner,

$$\mathcal{F}_\boldsymbol{\omega} \triangleq \left\{ \boldsymbol{\omega} \,\middle|\, \begin{array}{c} \mathbf{A}^T\boldsymbol{\omega} \leq -[\boldsymbol{\delta}^a]_-, \\ -[\boldsymbol{\delta}^b]_+ \leq \mathbf{B}^T\boldsymbol{\omega} \leq -[\boldsymbol{\delta}^b]_- \end{array} \right\}, \quad (13)$$

where $\boldsymbol{\delta}^a = \mathbf{A}^T\mathbf{x} + \mathbf{a}$, and $\boldsymbol{\delta}^b = \mathbf{B}^T\mathbf{x} + \mathbf{b}$.

$\nabla_\mathbf{x}\mathbf{f}(\mathbf{x})\boldsymbol{\xi}$ with $\boldsymbol{\xi} \geq \mathbf{0}_m, \mathbf{1}_m^T\boldsymbol{\xi} = 1$ may not guarantee that the incremented **x** remains in the feasible region. In ALMA the increment **ω** on **x** is a fraction $\nu \in (0,1]$ of that obtained by MGDA such that $\boldsymbol{\omega} \in \mathcal{F}_\boldsymbol{\omega}$. Furthermore, $\nu$ should be maximized so that the increment is as close to $\nabla_\mathbf{x}\mathbf{f}(\mathbf{x})\boldsymbol{\xi}$ as possible. This leads to the following bilevel problem,

$$\boldsymbol{\omega} = \nu \nabla_\mathbf{x}\mathbf{f}(\mathbf{x})\boldsymbol{\xi}, \quad (14a)$$

where,

$$\boldsymbol{\xi}, \nu = \underset{\substack{0 \leq \nu \leq 1, \nu \boldsymbol{\xi}^T \nabla_\mathbf{x}\mathbf{f} \in \mathcal{F}_\boldsymbol{\omega} \\ \boldsymbol{\xi} = \operatorname{argmin} \|\nabla_\mathbf{x}\mathbf{f}(\mathbf{x})\boldsymbol{\xi}\| \\ \boldsymbol{\xi} \geq \mathbf{0}, \mathbf{1}^T\boldsymbol{\xi} = 1}}{\operatorname{argmax}} \nu. \quad (14b)$$

This scheme is illustrated in Fig. 5 for a bi-objective CVOP. The shaded elliptical region represents $\mathcal{F}_\mathbf{x}$. The vector $\nabla_\mathbf{x}\mathbf{f}(\mathbf{x})\boldsymbol{\xi}$ (orange dotted arrow) is the perpendicular bisector of the shaded triangle shaped region whose sides are $\nabla_\mathbf{x} f_1(\mathbf{x})$ and $\nabla_\mathbf{x} f_2(\mathbf{x})$ (red, dotted arrows). The increment **ω** is the solid green arrow in Fig. 5.

Suppose **x** is infeasible – a situation that occurs commonly in exterior point algorithms such as the augmented Lagrangian method that ALMA incorporates. Unless the point **x** is sufficiently close to $\mathcal{F}_\mathbf{x}$, there may not exist any $\nu \in [0,1]$ such that $\boldsymbol{\omega} \in \mathcal{F}_\boldsymbol{\omega}$. As a result, ALMA does not increment **x** in the direction of the gradients until it is either inside $\mathcal{F}_\mathbf{x}$ or close enough to it. The point is still updated using the terms involving the dual variables. This is a desirable feature as it helps **x** move quicker towards the feasible region while allowing the dual variables acquire more consistent values.

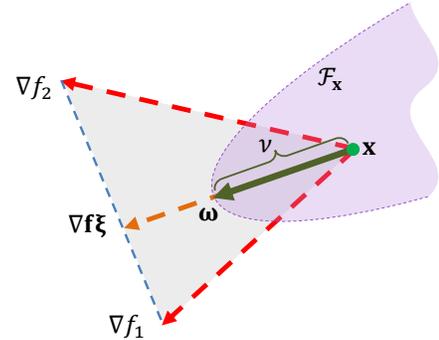

Fig. 5. Common feasible ascent direction

### C. DSO Level Pareto-Optimality

It is now shown that any locally Pareto optimal point obtained by ALMA is Pareto optimal.

**Theorem-2**. Jain's index $J(\cdot)$ is strictly quasiconcave in $\mathbb{R}^n$.

*Proof*: Let $\mathbf{x}, \mathbf{y} \in \mathbb{R}^n$ be two independent non-zero vectors such that $f(\mathbf{x}) \geq f(\mathbf{x} + \mathbf{y})$. For a function $f(\cdot)$ to be quasiconcave, $\mathbf{y}^T \nabla_\mathbf{x} f(\mathbf{x}) \geq 0$. Suppose **x, y** are such that $f(\mathbf{x}) \geq f(\mathbf{x} + \mathbf{y})$ and $\mathbf{y}^T \nabla_\mathbf{x} f(\mathbf{x}) = 0$. From the generalized mean value theorem, there must exist a $\rho \in [0,1]$, such that, $f(\mathbf{x} + \mathbf{y}) = f(\mathbf{x}) + \mathbf{y}^T \nabla_\mathbf{x} f(\mathbf{x}) + \frac{1}{2}\mathbf{y}^T \nabla_\mathbf{x}^2 f(\mathbf{x} + \rho(\mathbf{y} - \mathbf{x}))\mathbf{y}$. As $f(\mathbf{x}) \geq f(\mathbf{x} + \mathbf{y})$ and $\mathbf{y}^T \nabla_\mathbf{x} f(\mathbf{x}) = 0$, it must be true that the third term, $\mathbf{y}^T \nabla_\mathbf{x}^2 f(\mathbf{x} + \rho(\mathbf{y} - \mathbf{x}))\mathbf{y} \leq 0$. Letting $\rho = 0$, a sufficient condition for $f(\cdot)$ is that if $\mathbf{y}^T \nabla_\mathbf{x} f(\mathbf{x}) \geq 0$ for some $\mathbf{x}, \mathbf{y} \in \mathbb{R}^n$, then $\mathbf{y}^T \nabla_\mathbf{x}^2 f(\mathbf{x}) \mathbf{y} \leq 0$.

From (5) it can be shown that,
$$\nabla_\mathbf{x} J(\mathbf{x}) = 2\sqrt{J(\mathbf{x})} \left( \frac{1}{\|\mathbf{1}\|\|\mathbf{x}\|} - \sqrt{J(\mathbf{x})} \frac{\mathbf{x}}{\|\mathbf{x}\|^2} \right). \quad (15)$$

The Hessian of $J(\mathbf{x})$ can be obtained easily by differentiating the above expression,
$$\nabla_\mathbf{x}^2 J(\mathbf{x})\mathbf{y} = -\frac{2}{\|\mathbf{x}\|^2} \sqrt{J(\mathbf{x})} \left( \frac{\mathbf{1}\mathbf{x}^T}{\|\mathbf{1}\|\|\mathbf{x}\|} - \sqrt{J(\mathbf{x})} \frac{\mathbf{x}\mathbf{x}^T}{\|\mathbf{x}\|^2} \right)$$
$$- \frac{2}{\|\mathbf{x}\|^2} \left( \frac{-\mathbf{1}}{\|\mathbf{1}\|} + 2 \frac{\mathbf{x}}{\|\mathbf{x}\|} \sqrt{J(\mathbf{x})} \right) \left( \frac{\mathbf{1}^T}{\|\mathbf{1}\|} - \sqrt{J(\mathbf{x})} \frac{\mathbf{x}^T}{\|\mathbf{x}\|} \right)$$
$$- \frac{1}{\|\mathbf{x}\|^2} J(\mathbf{x}) \left( 2\mathbf{I} - \frac{\mathbf{x}\mathbf{x}^T}{\|\mathbf{x}\|^2} \right).$$

Rearranging terms and simplifying further using (15) leads to,
$$\nabla_\mathbf{x}^2 J(\mathbf{x}) = -\frac{1}{\|\mathbf{x}\|^2} \nabla_\mathbf{x} J(\mathbf{x}) \mathbf{x}^T$$
$$+ \left( \frac{1}{2J(\mathbf{x})} \nabla_\mathbf{x} J(\mathbf{x}) - \frac{\mathbf{x}}{\|\mathbf{x}\|^2} \right) \nabla_\mathbf{x}^T J(\mathbf{x})$$
$$- \frac{1}{\|\mathbf{x}\|^2} J(\mathbf{x}) \left( 2\mathbf{I} - \frac{\mathbf{x}\mathbf{x}^T}{\|\mathbf{x}\|^2} \right).$$

Using the above expression for the Hessian,
$$\mathbf{y}^T \nabla_\mathbf{x}^2 J(\mathbf{x}) \mathbf{y} = \frac{1}{\|\mathbf{x}\|^2} \mathbf{y}^T \nabla_\mathbf{x} J(\mathbf{x}) \mathbf{x}^T \mathbf{y}$$
$$+ \mathbf{y}^T \left( \frac{1}{2J(\mathbf{x})} \nabla_\mathbf{x} J(\mathbf{x}) - \frac{\mathbf{x}}{\|\mathbf{x}\|^2} \right) \nabla_\mathbf{x}^T J(\mathbf{x}) \mathbf{y}$$
$$- \frac{\mathbf{y}^T}{\|\mathbf{x}\|^2} J(\mathbf{x}) \left( 2\mathbf{I} - \frac{\mathbf{x}\mathbf{x}^T}{\|\mathbf{x}\|^2} \right) \mathbf{y}.$$

Under the assumption that $\mathbf{y}^T \nabla_\mathbf{x} J(\mathbf{x}) = 0$, the above equality can be simplified to,
$$\mathbf{y}^T \nabla_\mathbf{x}^2 J(\mathbf{x}) \mathbf{y} = -\frac{1}{\|\mathbf{x}\|^4} J(\mathbf{x}) (2 \|\mathbf{x}\|^2 \|\mathbf{y}\|^2 - (\mathbf{x}^T \mathbf{y})^2).$$

From the Cauchy-Schwarz inequality, as $\mathbf{x}$ and $\mathbf{y}$ are independent non-zero vectors, $|\mathbf{x}^T\mathbf{y}| < \|\mathbf{x}\|\|\mathbf{y}\|$ so that $\mathbf{y}^T \nabla_\mathbf{x}^2 J(\mathbf{x}) \mathbf{y} < 0$. This proves the strict quasiconcavity of $J(\mathbf{x})$. ∎

Theorem-3 below provides sufficient conditions for the convergence of ALMA towards Pareto optimal allocations.

***Theorem*-3**. If all scalar objectives $f_j(\cdot)$ of the vector function $\mathbf{f}(\cdot)$ are strictly quasiconcave, then any locally Pareto optimal point $\mathbf{x} \in \mathcal{F}_\mathbf{x}$ is globally Pareto optimal.

*Proof*: Let $\mathbf{x}$ be a locally Pareto optimal point. Thus there exists a $\sigma > 0$ such that $\mathbf{x}$ dominates every other feasible point in the ball $\mathcal{B}(\mathbf{x}, \sigma)$. Assume that, contrary to the statement of this theorem, $\mathbf{x}$ is not globally Pareto optimal. Under these circumstances we pick an arbitrary point $\mathbf{y} \in \mathcal{F}_\mathbf{x}$ with $\|\mathbf{x} - \mathbf{y}\| > \sigma$, such that $\mathbf{x} \not\succeq \mathbf{y}$. In other words, there is an objective $f_j(\cdot)$ such that $f_j(\mathbf{y}) > f_j(\mathbf{x})$. From (8), the feasible region $\mathcal{F}_\mathbf{x}$ is convex, so that for all $\theta \in [0,1]$, the point $\mathbf{z} = \theta\mathbf{x} + (1-\theta)\mathbf{y}$ must be feasible, i.e. $\mathbf{z} \in \mathcal{F}_\mathbf{x}$. From (9) the strict quasiconcavity of $f_j(\cdot)$ implies that $f_j(\mathbf{z}) > \min(f_j(\mathbf{x}), f_j(\mathbf{y}))$, i.e. $f_j(\mathbf{z}) > f_j(\mathbf{x})$. If $\theta$ is confined to the smaller interval, $(0, \|\mathbf{x} - \mathbf{y}\|^{-1}\sigma] \subset [0,1]$, then $\mathbf{z}$ lies inside the ball $\mathcal{B}(\mathbf{x}, \sigma)$. Since $\mathbf{z} \in \mathcal{B}(\mathbf{x}, \sigma)$, $\mathbf{x} \succeq \mathbf{z}$ so that $f_j(\mathbf{z}) \leq f_j(\mathbf{x})$, contradicting our previous observation that $f_j(\mathbf{z}) > f_j(\mathbf{x})$. ∎

Theorem-3 is of significance to the DSO CVOP defined in Section III.*D*. Since it has been assumed that all agents' utility functions are strictly quasiconcave, consequently, $\mathcal{U}_k(\cdot)$, which is their sum over disjoint arguments, is also strictly quasiconcave. By analogous reasoning, so is the welfare function $\mathcal{W}(\cdot)$. In Theorem-2, the Jain's index is shown to be strictly quasiconcave; therefore the fairness measure, $\mathcal{R}(\cdot)$ is also strictly quasiconcave. In other words, all scalar components of $\mathbf{\Omega}(\cdot)$ in (7) are strictly quasiconcave functions, whence by Theorem-3 the locally Pareto optimal point obtained by ALMA is Pareto optimal. More formal proofs of the theorem can be found in [44],[45].

### D. DSO Algorithm

The steps involved in the DSO algorithm are outlined below.

**until** (termination) **do**
1. Receive $\mathbf{c}$ from aggregators
2. $\mathbf{g} \leftarrow \nabla_\mathbf{p} \mathcal{R}(\mathbf{p})$
3. $\underline{\boldsymbol{\delta}}^V \leftarrow -\mathbf{C}^V \mathbf{p} + \mathbf{c}_l^V, \ \overline{\boldsymbol{\delta}}^V \leftarrow \mathbf{C}^V \mathbf{p} + \mathbf{c}_u^V$
   $\boldsymbol{\delta}^S \leftarrow \mathbf{C}^S \mathbf{p} + \mathbf{c}_0^S,$
   $\boldsymbol{\delta}^{P_0} \leftarrow \mathbf{C}^{P_0} \mathbf{p} + c_0^{P_0} - P_0, \quad \delta^B \leftarrow -\mathbf{c}^T\mathbf{p} + c_0 P_0$
4. $\mathbf{A} \leftarrow \begin{bmatrix} -\mathbf{C}^V \\ \mathbf{C}^V \\ \mathbf{C}^S \\ -\mathbf{c}^T \end{bmatrix} \quad \mathbf{a} \leftarrow \begin{bmatrix} -\mathbf{c}_l^V \\ -\mathbf{c}_u^V \\ -\mathbf{c}_0^S \\ -c_0 P_0 \end{bmatrix}$
   $\mathbf{B} \leftarrow \mathbf{C}^{P_0}, \quad \mathbf{b} \leftarrow c_0^{P_0} - P_0$
5. $\boldsymbol{\xi} \leftarrow \underset{\boldsymbol{\xi} \geq 0, \mathbf{1}^T\boldsymbol{\xi}=1}{\mathrm{argmin}} \|\nabla_\mathbf{x} \mathbf{f}(\mathbf{x})\boldsymbol{\xi}\|$
6. $\boldsymbol{\omega}' \leftarrow \nabla_\mathbf{x} \mathbf{f}(\mathbf{x})\boldsymbol{\xi}$
7. $\nu \leftarrow \underset{\substack{0 \leq \nu \leq \nu_{\max} \\ \nu\boldsymbol{\omega}' \in \mathcal{F}_\omega}}{\max} \nu$
8. $\boldsymbol{\omega} \leftarrow \nu \boldsymbol{\omega}'$
9. $\underline{\boldsymbol{\alpha}} \leftarrow [\underline{\boldsymbol{\alpha}} + \nu \eta_k^V \underline{\boldsymbol{\delta}}^V]_+, \quad \overline{\boldsymbol{\alpha}} \leftarrow [\overline{\boldsymbol{\alpha}} + \nu \eta_k^V \overline{\boldsymbol{\delta}}^V]_+$
   $\boldsymbol{\beta} \leftarrow [\boldsymbol{\beta} + \nu \eta_k^S \boldsymbol{\delta}^S]_+$
   $\lambda \leftarrow \lambda + \nu \eta_k^{P_0} \delta^{P_0}, \quad \gamma \leftarrow [\gamma + \nu \eta_k^B \delta^B]_+$
10. $\Delta\mathbf{p} \leftarrow \boldsymbol{\omega} + \mathbf{C}^{V^T}\underline{\boldsymbol{\alpha}} - \mathbf{C}^{V^T}\overline{\boldsymbol{\alpha}} - \mathbf{C}^{S^T}\boldsymbol{\beta} - \lambda \mathbf{C}^{P_0} + \gamma \mathbf{c}$
    $\mathbf{p} \leftarrow \mathbf{p} + \eta_k^P \Delta\mathbf{p}$
11. Send $\mathbf{p}$ to aggregators
12. $k \leftarrow k + 1$

**end**

The DSO algorithm is a specific implementation of ALMA for energy allocation in distribution systems. In step 1, the aggregator receives the unit cost $\mathbf{c}$ from aggregator auctions, which is equal to $\nabla_\mathbf{p} \mathcal{W}(\mathbf{p})$ (Theorem-1). The other gradient, $\nabla_\mathbf{p} \mathcal{R}(\mathbf{p})$ is computed in step 2 where $\mathbf{p}$ is the value from the previous iteration. Using the constraint gaps $\boldsymbol{\delta}^*$ (step 3) the quantities involved in $\mathcal{F}_\mathbf{x}$ in (8) are determined (step 4). Steps 5–8 implement (12) as follows. Steps 5 and 6 implement ascent direction as in MGDA. This is scaled by the factor $\nu$ so that increments do not produce infeasible solutions. The dual variables are incremented (step 9) in accordance with augmented Lagrangian method (see [14] for details), following which the energy allocation is incremented (step 10), and returned to the aggregator (step 11) for the next round of aggregator auctions until convergence towards a locally Pareto optimum, which Theorem-3 shows to be Pareto optimal.

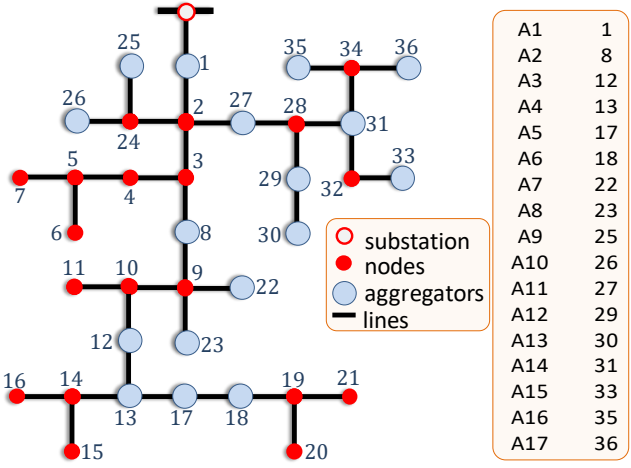

| | |
|---|---|
| A1 | 1 |
| A2 | 8 |
| A3 | 12 |
| A4 | 13 |
| A5 | 17 |
| A6 | 18 |
| A7 | 22 |
| A8 | 23 |
| A9 | 25 |
| A10 | 26 |
| A11 | 27 |
| A12 | 29 |
| A13 | 30 |
| A14 | 31 |
| A15 | 33 |
| A16 | 35 |
| A17 | 36 |

Fig. 6. IEEE 37-bus system used as simulation platform

## V. SIMULATION RESULTS

The proposed approach use was implemented on a modified IEEE 37-bus system as shown in Fig. 6. Nodes containing the 17 aggregators appear as larger blue circles, whereas the remaining nodes are red filled circles. For convenience, the aggregators are indexed separately (inset in Fig. 6) The number of agents in each aggregator was generated randomly between $G_k = 9$ and $G_k = 25$, with aggregators A4, A6, A10, A12 having a higher number of prosumers. Some agents were equipped with some PV generation ($g_k^i > 0$). The agent parameters, $a_k^i$, $b_k^i$, and $g_k^i$ (see Fig. 2) were generated randomly. All simulations were performed in MATLAB.

In order to see the effect of fairness, two simulations were done. The algorithm in [7] was implemented to obtain the efficient solution without any fairness. Following this, the DSO algorithm was implemented. Fig. 7 compares the results of both simulations. The blue vertical bars are the power allocations of the aggregators ($p_k$) that were computed from the simulations without fairness. The aggregators' power allocations are shown as vertical bars that are colored blue (without fairness) and yellow (with fairness). The solid lines (without fairness) and dotted lines (with fairness) in the figure show the unit costs. The quantities with fairness are superscripted with asterisks (*).

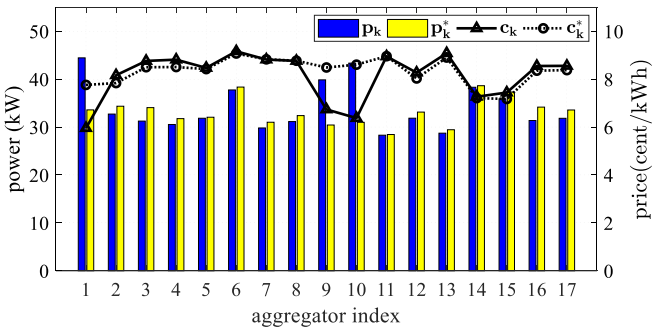

Fig. 7. Aggregator power allocations and unit costs

From Fig. 7 it can be seen that aggregators A1, A9, A10 received more power at lower unit cost in the absence of fairness. As can be seen in in Fig. 6, these aggregators are positioned closer to the substation node (red circle with white interior). In contrast, aggregators A3, A12, A16, and A17 which are further away, experience higher unit cost and lower power allocation. The allocations obtained with fairness show how utilizing the Jain's index helps in mitigating this adverse effect. The fairness objective causes aggregators to be charged in a more equitable manner.

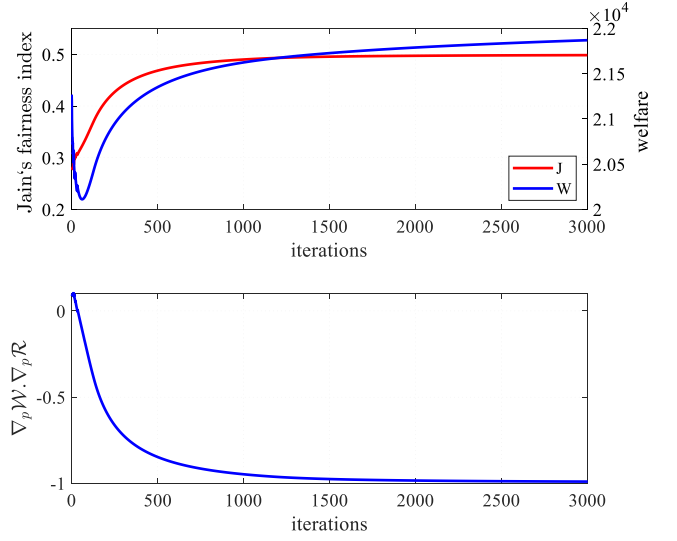

Fig. 8. Welfare and Jain's index (top), inner product of gradients of welfare and fairness (bottom) vs. iteration

Fig. 8 shows the progress of the algorithm with iteration. Fig 8 (top) shows how the inner product $\langle \mathcal{W}(\mathbf{p}), \mathcal{R}(\mathbf{p}) \rangle$ converges towards $-1$ at the Pareto front. The steadily increasing welfare (blue line) and fairness (red line) are shown in Fig. 8 (bottom). Note that during the initial stages of the algorithm, the solution would be infeasible, explaining the initial fluctuations. Although 3,000 iterations were allowed for convergence, with very small random initialization of $\mathbf{p}$ and zero initial duals, with higher initial values, the algorithm would require as little as 500 iterations to converge, which is not much more than in [11] despite the presence of a vector objective.

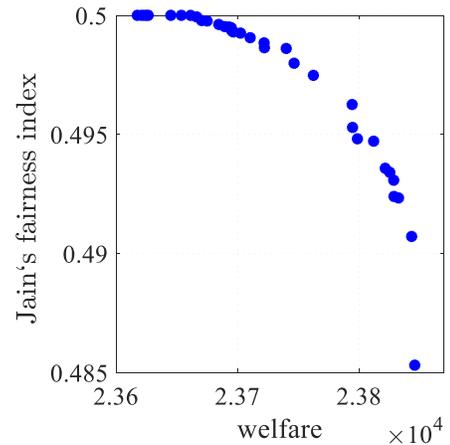

Fig. 9. Jain's index vs. welfare

Lastly, the tradeoff between welfare and fairness was investigated. Multiple simulations were carried out with random initialization of the primal and dual variables. Fig. 9 depicts the resulting Pareto front.

## VI. Conclusions

This research proposes a general-purpose approach for constrained vector optimization. The analytical treatment, albeit informal, shows that Pareto optimal solutions can be obtained as long as the objectives are strictly quasiconcave. This is an improvement over previous approaches, e.g. in, [4], [6],[7],[13],[38],[8], which routinely make more restrictive assumption of concave utilities. This is of significance as general econometric theory does not support prior assumptions of concavity, instead treating utilities justifiably as quasiconcave functions [47],[48],[49]. Efficiency-fairness tradeoff is a crucial issue in resource allocation [39],[50], [51],[52] with a significant amount of research using Jain's index (cf. [14],[53],[54]). In establishing the quasiconcavity of Jain's index, this research provides a theoretical justification for the application of vector optimization algorithms such as ALMA for tradeoff allocations.

It should be noted here that this research relies significantly on recent research. In particular, ALMA obtains the gradient in the same direction as in MGDA. The dual variables are incremented by the DSO algorithm in the same manner as in [14]. The aggregator auction (Section III.A) is a modification of those in [7],[38]. Additionally, there are a few limitations of this research that are outlined below.

It has been assumed that the feasibility constraints in (8) were linear equality and inequality constraints. Although supported by (3) for the energy grid, this assumption is over-simplistic for other applications. One possible improvement would be to linearize any nonlinear constraints at intermittent stages of the optimization algorithm. However, the effectiveness of ALMA in such situations needs to be further investigated.

The Pareto front in Fig. 9 was obtained by randomly initializing the starting point. However, this approach highlights that ALMA can converge to any Pareto-optimal solution. Although the present simulation results indicated that any additional gain in efficiency was accompanied by a sharp drop in fairness, this may not necessarily be the case in other grids, and therefore is a potential limitation in ALMA. One option to exert more influence on the Pareto-optimal output allocation of ALMA, the authors suggest using MGDA's approach to deal with opposing objectives, which suggests the use of relaxation algorithms to converge towards generalized Nash equilibrium. Alternately, only one of the scalar functions in (7) may be used as a scalar objective, while imposing bounds either on the other objective or on both, as additional constraints. A similar constrained method has been used in [37]. Both approaches while remaining quite out of scope in this study, are worthwhile directions for further investigation.

In the absence of any secondary stage to navigate the Pareto front until a user-specified suitable point is reached, how useful is ALMA for use in the energy grid? Fortunately, in day-ahead planning where scheduling is usually done in an hourly manner, the allocation $\mathbf{p}_t$ during any hourly time interval $t$ would not differ significantly from that of the previous interval $\mathbf{p}_{t-1}$. Initializing $\mathbf{p}_t$ to the previous $\mathbf{p}_{t-1}$, which is not only be feasible but also located very close to the desirable region in the Pareto front, would allow ALMA to converge to a Pareto optimal allocation at least an order of magnitude faster than what Fig 8 suggests. ALMA can be used in a similar fashion during real time operation, when actual user demands deviate from their forecasts. Barring unforeseen weather changes, as such deviations are usually very small, the planned value of $\mathbf{p}_t$ can readily be used as the initial point. Alternately, historical values from the DSO's database can also be adopted for initialization during weather related exigencies. Put together, these reasons largely obviate the need for ALMA to be equipped to move along the Pareto front.

The step sizes $\eta_k^X$ of each dual variable $X$ was obtained by the DSO algorithm as in [14]. However, energy allocations were incremented with $\eta_k^P$ being kept proportional to $(1 + \langle \mathcal{W}(\mathbf{p}), \mathcal{R}(\mathbf{p}) \rangle)$. This method of stepwise updates made ALMA apply increasingly smaller increments as it approached the Pareto front. Although simulations in this research indicated its effectiveness, theoretical support for such a modification is lacking. The authors intend to extend this technique for more than two objectives, and to formally establish convergence limits with step sizes fashioned in this manner.

The proposed approach should be compared with novel algorithms for CVOP that were published recently in [35],[36], both of which appeared during a later phase of this research. In a similar manner, the effectiveness of ALMA with more than only two objectives, should be investigated in future research. The simulation results reported here serve as a proof-of-concept for a more general-purpose approach for large-scale constrained vector optimization.

## VII. Acknowledgment

This work was supported by the National Science Foundation-CPS under Grant CNS-1544705.